\documentstyle[eqsecnum,aps,graphicx]{revtex}

\begin{document}
\draft
\title{Quantum Isotropization of the Universe}
\author{N. Pinto-Neto\thanks{%
e-mail address: {\tt nelsonpn@lafex.cbpf.br}}, A. F. Velasco\thanks{%
e-mail address: {\tt velasco@lca1.drp.cbpf.br}}}
\address{Centro Brasileiro de Pesquisas F\'{\i}sicas, \\
Rua Dr.\ Xavier Sigaud 150, Urca 22290-180 -- Rio de Janeiro, RJ -- Brazil}
\author{and R. Colistete Jr.\thanks{
e-mail address: {\tt coliste@ccr.jussieu.fr}}}
\address{Laboratoire de Gravitation et Cosmologie Relativistes,
Universit\'e Pierre et Marie Curie, \\
Tour 22, 4\`eme \'etage, Bo\^{\i}te 142, 4 place Jussieu,
75252 Paris Cedex 05, France}
\date{\today}
\maketitle

\begin{abstract}
We consider minisuperspace models constituted of Bianchi I geometries with a
free massless scalar field. The classical solutions are always singular
(with the trivial exception of flat space-time), and always anisotropic once
they begin anisotropic. When quantizing the system, we obtain the
Wheeler-DeWitt equation as a four-dimensional massless Klein-Gordon
equation. We show that there are plenty of quantum states whose
corresponding bohmian trajectories may be non-singular and/or presenting
large isotropic phases, even if they begin anisotropic, due to quantum
gravitational effects. As a specific example, we exhibit field plots of
bohmian trajectories for the case of gaussian superpositions of plane wave
solutions of the Wheeler-DeWitt equation which have those properties.
These conclusions are valid even in the absence of the scalar field.
\vspace{0.7cm}

\end{abstract}


\section{Introduction}

Two of the main questions one might ask in cosmology are:

1) Is the Universe eternal or it had a beginning, and in the last case, was
this beginning given by an initial singularity?

2) Why the Universe we live in is remarkably homogeneous and isotropic, with
very small deviations from this highly symmetric state?

The answer given by classical General Relativity (GR) to the first question,
indicated by the singularity theorems \cite{he}, asserts that probably the
Universe had a singular beginning. As singularities are out of the scope of
any physical theory, this answer invalidates any description of the very
beginning of the Universe in physical terms. One might think that GR and/or
any other matter field theory must be changed under the extreme situations
of very high energy density and curvature near the singularity, rendering
the physical assumptions of the singularity theorems invalid near this
point. One good point of view (which is not the only one) is to think that
quantum gravitational effects become important under these extreme
conditions. We should then construct a quantum theory of gravitation and
apply it to cosmology. For instance, in the euclidean quantum gravity
approach \cite{haw} to quantum cosmology, a second answer to the first
question comes out: the Universe may have had a non-singular birth given by
the beginning of time through a change of signature.

In the same way, the naive answer of GR and the standard cosmological
scenario to the second question is not at all satisfactory: the reason for
the Universe be highly homogeneous and isotropic is a matter of initial
conditions. However, solutions of Einstein's equations with this symmetry
are of measure zero; so, why the Universe is not inhomogeneous and/or
anisotropic? Inflation \cite{gut,kol} is an idea that tries to explain this
fact. Nevertheless, in order for inflation to happen, some special initial
conditions are still necessary, although much more less stringent than in
the case without inflation. Once again, quantum cosmology can help in this
matter by providing the physical reasons for having the initial conditions
for inflation.

Recently, many papers have been devoted to the application of the Bohm-de
Broglie interpretation of quantum mechanics \cite{boh,hol} to quantum
cosmology \cite{vink,kow,sht,val,bola1,bola2,bola3,fab,fab2,nel}. One of the
aims of these papers was to try to give an answer to the first question
within a new perspective in quantum cosmology. In the framework of the
Bohm-de Broglie interpretation of quantum mechanics applied to
minisuperspace models, it was shown in References \cite
{bola1,bola2,bola3,fab,fab2} that quantum gravitational effects may indeed
become important under extreme situations of very high energy density and
curvature, and avoid the initial singularity of the classical models. The
solutions that come out are then non-singular (eternal) cosmological models,
but with a very hot phase, tending to the usual classical solutions when the
energy scales become smaller than the Planck scale. Of course this is a much
better answer than the one given by classical GR.

The aim of this paper, as a natural sequence of the references cited above,
is to address the second question from the point of view of the Bohm-de
Broglie interpretation of quantum cosmology. We want to investigate if
quantum mechanical effects can isotropize an anisotropic cosmological model
yielding an alternative mechanism for the isotropization of the Universe.
Our strategy is to take classical anisotropic models which never
isotropizes, quantize them, and examine if quantum effects can at least
create isotropic phases. As we will see, quantum effects can not only create
isotropic phases but they can also isotropize the models forever, for a
large variety of quantum states. Other authors have also studied this
problem, mainly taking Bianchi IX models, adopting other interpretations of
quantum cosmology, and they arrive at similar results \cite
{mos,ams,ber,vil,mon}.

The mathematical models we take are Bianchi I models with and without
matter. For the case without matter, one of the anisotropic degrees of
freedom is suppressed and we end with a two-dimensional minisuperspace model.
The classical solutions of this case are all singular and always
anisotropic, with the trivial exception of flat space-time. When quantizing
this system we find that the solutions of the Wheeler-DeWitt can be written
in terms of superpositions of plane wave solutions. Taking gaussian
superpositions we show that there are bohmian trajectories representing
universes expanding from an initial singularity and others representing
bouncing non-singular universes, both with isotropic phases during the
course of their evolutions. There are also bohmian non-singular periodic
solutions with periodic isotropic phases and non-singular bouncing solutions
which are never isotropic. For the case with matter, we take a minimally
coupled scalar field with arbitrary coupling constant $\omega$. The
minisuperspace is four-dimensional. The classical solutions are again
singular, and once they begin anisotropic they continue to be anisotropic
forever. When quantizing the system, we arrive at a Wheeler-DeWitt equation
which corresponds to a massless Klein-Gordon equation in four dimensions. We
show that there are plenty of spherical wave solutions whose bohmian
trajectories represent expanding universes that isotropizes permanently
after some period, many of them which are also non-singular. gaussian
superpositions also present this feature.

The paper is organized as follows: in the next section we present the
minisuperspace models which we will investigate and their classical
solutions. In section III we quantize these models, and obtain their
corresponding Wheeler-DeWitt equations and general quantum solutions. In
section IV we present the results concerning the quantum isotropization of
those solutions through the investigation of the bohmian trajectories we
obtain from them. We end up in section V with conclusions and discussions.

\section{The classical minisuperspace model}

Let us take the lagrangian

\begin{equation}
{\it L} = \sqrt{-g}e^{-\phi}\biggr(R - w \phi_{;\rho}\phi^{;\rho}\biggl) \;.
\end{equation}
For $w=-1$ we have effective string theory without the Kalb-Ramond field.
For $w=-3/2$ we have a conformally coupled scalar field. Performing the
conformal transformation $g_{\mu\nu} = e^{\phi}{\bar g}_{\mu\nu}$ we obtain

\begin{equation}  \label{lg2}
{\it L} = \sqrt{-g}\biggr[R - C_w \phi_{;\rho} \phi^{;\rho}\biggl] \;,
\end{equation}
where the bars have been omitted, and $C_w \equiv (\omega + \frac{3}{2})$.
The cases of interest correspond to $C_w > 0$.

The gravitational part of the minisuperspace model we study in this paper is
given by the homogeneous and anisotropic Bianchi I line element
\begin{eqnarray} \label{bia}
ds^{2} &=&-N^{2}(t)dt^{2}+\exp [2\beta _{0}(t)+2\beta _{+}(t)+2\sqrt{3}\beta
_{-}(t)]\;dx^{2}+ \nonumber \\
&&\exp [2\beta _{0}(t)+2\beta _{+}(t)-2\sqrt{3}\beta _{-}(t)]\;dy^{2}+
\exp [2\beta _{0}(t)-4\beta _{+}(t)]\;dz^{2} \;.
\end{eqnarray}
This line element will be isotropic if and only if $\beta _{+}(t)$ and $%
\beta _{-}(t)$ are constants \cite{he}. Inserting Equation (\ref{bia}) into
the action $S=\int {{\it L\,}d^{4}x}$, supposing that the scalar field $\phi $
depends only on time, discarding surface terms, and performing a Legendre
transformation, we obtain the following minisuperspace classical hamiltonian
\begin{equation} \label{hbiaf}
H=\frac{N}{24\exp {(3\beta _{0})}}(-p_{0}^{2}+p_{+}^{2}+p_{-}^{2}+p_{\phi
}^{2}) \;,
\end{equation}
where $(p_{0},p_{+},p_{-},p_{\phi })$ are canonically conjugate to $(\beta
_{0},\beta _{+},\beta _{-},\phi )$, respectively, and we made the trivial
redefinition $\phi \rightarrow \sqrt{C_{w}/6}\;\;\phi $.

We can write this hamiltonian in a compact form by defining $y^{\mu} =
(\beta _0, \beta _+, \beta _-, \phi)$ and their canonical momenta $p_{\mu} =
(p_0, p_+, p_-, p_{\phi})$, obtaining
\begin{equation}  \label{ham}
H = \frac{N}{24 \exp{(3y^0)}}\eta ^{\mu\nu}p_{\mu}p_{\nu} \;,
\end{equation}
where $\eta ^{\mu\nu}$ is the Minkowski metric with signature $(-+++)$. The
equations of motion are the constraint equation obtained by varying the
hamiltonian with respect to the lapse function $N$

\begin{equation}  \label{hbia1f}
{\cal H} \equiv \eta ^{\mu\nu}p_{\mu}p_{\nu} = 0 \;,
\end{equation}
and the Hamilton's equations

\begin{equation}  \label{hbia2f}
\dot{y}^{\mu} = \frac{\partial{\cal H}}{\partial p_{\mu}} = \frac{N}{12 \exp{%
(3y_0)}}\eta ^{\mu\nu}p_{\nu} \;,
\end{equation}

\begin{equation}  \label{hbia3f}
\dot{p}_{\mu} = -\frac{\partial{\cal H}}{\partial y^{\mu}} = 0 \;.
\end{equation}
The solution to these equations in the gauge $N=12\exp (3 y_0)$ is

\begin{equation}  \label{solc}
y^{\mu} = \eta ^{\mu\nu}p_{\nu}t + C^{\mu} \;,
\end{equation}
where the momenta $p_{\nu}$ are constants due to the equations of motion and
the $C^{\mu}$ are integration constants. We can see that the only way to
obtain isotropy in these solutions is by making $p_{1}=p_{+}=0$ and $%
p_{2}=p_{-}=0$, which yields solutions that are always isotropic, the usual
Friedmann-Robertson-Walker (FRW) solutions with a scalar field. Hence, there
is no anisotropic solution in this model which can classically becomes
isotropic during the course of its evolution. Once anisotropic, always
anisotropic. If we suppress the $\phi$ degree of freedom, the unique
isotropic solution is flat space-time because in this case the constraint (%
\ref{hbia1f}) enforces $p_0 =0$.

To discuss the appearance of singularities, we need the Weyl square tensor $%
W^{2}\equiv W^{\alpha \beta \mu \nu }W_{\alpha \beta \mu \nu }$. It reads
\begin{equation} \label{w2}
W^{2}=\frac{1}{432}e^{-12\beta
_{0}}(2p_{0}p_{+}^{3}-6p_{0}p_{-}^{2}p_{+}+p_{-}^{4}+2p_{+}^{2}p_{-}^{2}+
p_{+}^{4}+p_{0}^{2}p_{+}^{2}+p_{0}^{2}p_{-}^{2}) \;.
\end{equation}
Hence, the Weyl square tensor is proportional to $\exp {(-12\beta _{0})}$
because the $p$'s are constants (see Equations (\ref{hbia3f})), and the
singularity is at $t=-\infty $. The classical singularity can be avoided only
if we set $p_{0}=0$. But then, due to Equation (\ref{hbia1f}), we would also
have $p_{i}=0$, which corresponds to the trivial case of flat space-time.
Therefore, the unique classical solution which is non-singular is the trivial
flat space-time solution.

\section{Quantization of the classical model}

The Dirac quantization procedure yields the Wheeler-DeWitt equation through
the imposition of the condition
\begin{equation}  \label{nense}
\hat{{\cal H}} \Psi = 0 \;,
\end{equation}
on the quantum states, with $\hat{{\cal H}}$ defined as in Equation
(\ref{hbia1f}) (we are assuming the covariant factor ordering) using the
substitutions
\begin{equation}
p_{\mu} \rightarrow - i\frac{\partial }{\partial y^{\mu}} \;.
\end{equation}
Equation (\ref{nense}) reads
\begin{equation}  \label{wdw}
\eta ^{\mu\nu}\frac{\partial ^2}{\partial y^{\mu} y^{\nu}} \Psi (y^{\mu}) =0 \;.
\end{equation}

\subsection{The empty model}

In a first and simple example, we will freeze two degrees of freedom: the
matter degree of freedom $y^3=\phi=0$, and one of the anisotropic degrees of
freedom, $y^{2}=\beta_{-}=0$. We obtain a two-dimensional Klein-Gordon
equation whose general solution is
\begin{equation}
\Psi(\beta _0,\beta _+) = \int{\{F(k)\exp[ik(\beta _0 + \beta _+)] +G(k)\exp[%
ik(\beta _0 - \beta _+)]\}dk} \;,
\end{equation}
with $F(k)$ and $G(k)$ arbitrary functions of $k$. Let us take gaussian
superpositions

\begin{equation}  \label{gauss1}
F(k) = G(-k) = \exp\biggr[-\frac{(k-d)^2}{\sigma ^2}\biggl] \;,
\end{equation}
where $\sigma$ and $d$ are constants. In this case the wave function reads
\cite{gra}

\begin{equation}  \label{psi1}
\Psi _1 = \sigma \sqrt{\pi} \biggl\{ \exp\biggl[-\frac{(\beta _0 + \beta
_+)^2 \sigma ^2}{4}\biggr] \exp[id(\beta _0 + \beta _+)] + \exp\biggl[-\frac{%
(\beta _0 - \beta _+)^2 \sigma ^2}{4}\biggr] \exp[-id(\beta _0 - \beta _+)]%
\biggr\} \;.
\end{equation}

For the case where

\begin{equation}  \label{gauss2}
F(k) = G(k) = \exp\biggr[-\frac{(k+d)^2}{\sigma ^2}\biggl] \;,
\end{equation}
the wave function turns out to be

\begin{equation}  \label{psi2}
\Psi _2 = \sigma \sqrt{\pi} \biggl\{ \exp\biggl[-\frac{(\beta _0 + \beta
_+)^2 \sigma ^2}{4}\biggr] \exp[-id(\beta _0 + \beta _+)] + \exp\biggl[-%
\frac{(\beta _0 - \beta _+)^2 \sigma ^2}{4}\biggr] \exp[-id(\beta _0 - \beta
_+)]\biggr\} \;.
\end{equation}

In the next section we will calculate the bohmian trajectories corresponding
to the solutions (\ref{psi1},\ref{psi2}) and see if there are isotropic
phases in these quantum models.

\subsection{The scalar field model}

In the general case of the four-dimensional minisuperspace model, we will
investigate spherical-wave solutions of Equation (\ref{wdw}). They read

\begin{equation}  \label{psi3}
\Psi_3 = \frac{1}{y}\biggl[ f(y^0 + y) + g(y^0 - y)\biggr] \;,
\end{equation}
where $y\equiv \sqrt{\sum _{i=1}^{3} (y^i)^2}$.

One particular example is the gaussian superposition of plane wave solutions
of Equation (\ref{wdw}),
\begin{equation} \label{naosei}
\Psi _{4}(y^{\mu })=\int {\{F(\vec{k})\exp [i(|\vec{k}|y^{0}+\vec{k}.\vec{y}%
)]+G(\vec{k})\exp [i(|\vec{k}|y^{0}-\vec{k}.\vec{y})]\}d^{3}k} \;,
\end{equation}
where $\vec{k}\equiv (k_{1},k_{2},k_{3})$, $\vec{y}\equiv
(y^{1},y^{2},y^{3}) $, $|\vec{k}|\equiv \sqrt{\sum_{i=1}^{3}(k_{i})^{2}}$,
with $F(\vec{k})$ and $G(\vec{k})$ given by
\begin{equation} \label{gauss4}
F(\vec{k})=G(\vec{k})=\exp \biggr[-\frac{(|\vec{k}|-d)^{2}}{\sigma ^{2}}%
\biggl] \;.
\end{equation}
After performing the integration in Equation (\ref{naosei}) using spherical
coordinates we obtain \cite{gra}

\begin{eqnarray}  \label{psi4}
\Psi _{4}(y^{0},y) &=&\frac{i\pi ^{3/2}}{y}\biggl\{[2d\sigma
+i(y^{0}-y)\sigma ^{3}]\exp \biggl[-\frac{(y^{0}-y)^{2}\sigma ^{2}}{4}\biggr]%
\exp [id(y^{0}-y)] \biggl[1+\Phi \biggl(\frac{d}{\sigma }+i(y^{0}-y)%
\frac{\sigma }{2}\biggr)\biggr]  \nonumber  \\
&&-[2d\sigma +i(y^{0}+y)\sigma ^{3}]\exp \biggl[-\frac{(y^{0}+y)^{2}\sigma%
^{2}}{4}\biggr]\exp [id(y^{0}+y)] \biggl[1+\Phi \biggl(\frac{d}{\sigma }+%
i(y^{0}+y)\frac{\sigma }{2}\biggr)\biggr]\biggr\} \;,
\end{eqnarray}
where $\Phi (x)\equiv (2/\sqrt{{\pi }})\int_{0}^{x}\exp (-t^{2})dt$ is the
probability integral. The wave function $\Psi _{4}$ is a spherical solution
with the form of Equation (\ref{psi3}) with $g=-f$. In order to simplify $\Psi
_{4}$, we will take the limit $\sigma ^{2}>>d$ and $(y^{0} \pm y)\sigma>>1$ in
Equation (\ref{psi4}) yielding \cite{gra}
\begin{equation} \label{fapprox}
f(z)\approx -\frac{16\pi d}{\sigma ^{2}z^{3}} +
            i 2 \pi\biggl(\frac{2}{z^{2}} + \sigma ^{2}\biggr) \;.
\end{equation}

\section{The quantum bohmian trajectories and results}

In this section, we will apply the rules of the Bohm-de Broglie
interpretation to the wave functions we have obtained in the previous
section. We first summarize these rules for the case of homogeneous
minisuperspace models. In this case, the general minisuperspace
Wheeler-DeWitt equation is
\begin{equation}  \label{bsc}
{\cal H}[{\hat{p}}_{\alpha}(t), {\hat{q}}^{\alpha}(t)] \Psi (q) = 0 \;,
\end{equation}
where $p_{\alpha}(t)$ and $q^{\alpha}(t)$ represent the homogeneous degrees
of freedom coming from the gravitational and matter fields.
Writing $\Psi = R \exp (iS/\hbar)$, and substituting it into (\ref{bsc}), we
obtain the equation
\begin{equation}  \label{hoqg}
\frac{1}{2}f^{\alpha\beta}(q^{\mu})\frac{\partial S}{\partial q^{\alpha}}
\frac{\partial S}{\partial q^{\beta}}+ U(q^{\mu}) + Q(q^{\mu}) = 0 \;,
\end{equation}
where
\begin{equation}  \label{hqgqp}
Q(q^{\mu}) = -\frac{1}{2R} f^{\alpha\beta}\frac{\partial ^2 R} {\partial
q^{\alpha} \partial q^{\beta}} \;.
\end{equation}
The quantities $f^{\alpha\beta}(q^{\mu})$ and $U(q^{\mu})$ are the
minisuperspace particularizations of the DeWitt metric \cite{dew}, and of the
scalar curvature density of the space-like hypersurfaces together with matter
potential energies, respectively. The causal interpretation applied to
quantum cosmology states that the trajectories $q^{\alpha}(t)$ are real,
independently of any observations. Equation (\ref{hoqg}) is the Hamilton-Jacobi
equation for them, which is the classical one amended with the quantum
potential term (\ref{hqgqp}), responsible for the quantum effects. This
suggests to define
\begin{equation}  \label{h}
p_{\alpha} = \frac{\partial S}{\partial q^{\alpha}} \;,
\end{equation}
where the momenta are related to the velocities in the usual way
\begin{equation}  \label{h2}
p_{\alpha} = f_{\alpha\beta}\frac{1}{N}\frac{\partial q^{\beta}}{\partial t} \;,
\end{equation}
and $f_{\alpha\beta}$ is the inverse of $f^{\alpha\beta}$. To obtain the
quantum trajectories, we have to solve the following system of first order
differential equations, called the guidance relations:
\begin{equation}  \label{h3}
\frac{\partial S(q^{\gamma})}{\partial q^{\alpha}} = f_{\alpha\beta}\frac{1}{%
N}\frac{\partial q^{\beta}}{\partial t} \;.
\end{equation}
Equations (\ref{h3}) are invariant under time reparametrization. Hence, even
at the quantum level, different choices of $N(t)$ yield the same space-time
geometry for a given non-classical solution $q^{\alpha}(t)$. There is no
problem of time in the causal interpretation of minisuperspace quantum
cosmology\footnote{%
This is not the case, however, for the full superspace (see Reference
\cite{nel}).}.

For the minisuperspace we are investigating, the guidance relations in the
gauge $N=12\exp (3 y_0)$ are (see Equations (\ref{hbia2f}))

\begin{equation}  \label{gui}
p_{\mu} = \frac{\partial S}{\partial y^{\mu}} = \eta _{\mu\nu}{\dot{y}}%
^{\nu} \;,
\end{equation}
where $S$ is the phase of the wave function.

\subsection{The empty model}

Taking the phase of the solution (\ref{psi1}), and inserting it into Equations
(\ref{gui}), remembering that $y^{2}=\beta_{-}$ and $y^{3}=\phi$ have been
suppressed, we get the following system of planar equations:

\begin{equation}  \label{dbeta0dtpsi1}
\dot{\beta} _0 = \frac{\beta _+ \sigma ^2 \sin (2d\beta _0 ) + 2d\sinh
(\beta _0 \beta _+ \sigma ^2)} {2[ \cos (2d\beta _0 ) + \cosh (\beta _0
\beta _+ \sigma ^2)]} \;,
\end{equation}
\begin{equation}  \label{dbetamaisdtpsi1}
\dot{\beta} _+ = \frac{2d \cos(2d\beta _0 ) + 2d\cosh (\beta _0
 \beta _+ \sigma ^2) -\beta _0 \sigma ^2\sin (2d\beta _0 ) }
{2[ \cos (2d\beta _0 ) + \cosh (\beta _0 \beta _+ \sigma ^2)]} \;.
\end{equation}
The line $\beta _0 = 0$ divides configuration space in two symmetric
regions. The line $\beta _+ = 0$ contains all singular points of this
system, which are nodes and centers. The nodes appear when the denominator
of the above equations, which is proportional to the absolute value of the wave
function, is zero. No trajectory can pass through them. They happen
when $\beta _+ = 0$ and $\cos(2d\beta _0 ) = -1$, or $\beta _0 = (2n+1)\pi
/2d $, $n$ an integer, with separation $\pi/d$. The center points appear
when the numerators are zero. They are given by $\beta _+ = 0$ and $\beta _0
= 2 d\,[{\rm cotan} (d\beta _0 )]/\sigma ^2$. They are intercalated with the
node points. As $\mid \beta _0 \mid \rightarrow \infty$ these points tend to
$n\pi/d$, and their separations cannot exceed $\pi/d$. As one can see from
Equations (\ref{dbeta0dtpsi1}) and (\ref{dbetamaisdtpsi1}), the classical
solutions (\ref{solc}) are recovered when $\mid \beta _0 \mid \rightarrow
\infty$ and $\mid \beta _+ \mid \rightarrow \infty$, the $\beta$'s become
proportional to $t$. The quantum potential given in Equation (\ref{hqgqp}),
which now reads,

\begin{eqnarray}  \label{qp1}
Q(\beta _{0},\beta _{+}) &=&-\frac{1}{2R}\biggl(-\frac{\partial ^{2}R}{%
\partial \beta _{0}^{2}}+\frac{\partial ^{2}R}{\partial \beta _{+}^{2}}%
\biggr)  \nonumber  \\
&=&\frac{[\beta _{+}\sigma ^{2}\sin (2d\beta _{0})+2d\sinh (\beta _{0}\beta
_{+}\sigma ^{2})]^{2}-[\beta _{0}\sigma ^{2}\sin (2d\beta _{0})-2d\cos
(2d\beta _{0})-2d\cosh (\beta _{0}\beta _{+}\sigma ^{2})]^{2}}{8[\cos
(2d\beta _{0})+\cosh (\beta _{0}\beta _{+}\sigma ^{2})]^{2}} \;,
\end{eqnarray}
becomes negligible in these limits.

\begin{figure}
\begin{center}
\includegraphics[trim=0in -0.1in 0in 0in]{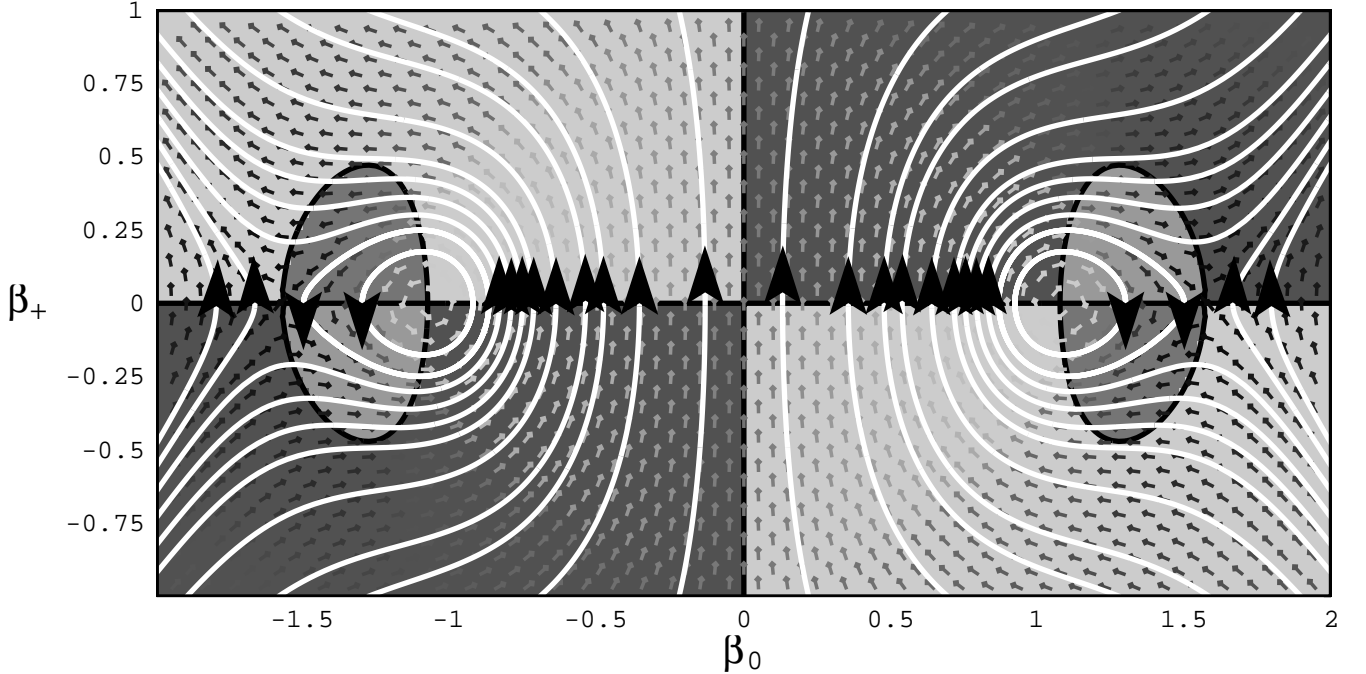}
\caption{
Field plot of the system of planar equations (\ref{dbeta0dtpsi1}-%
\ref{dbetamaisdtpsi1}) for $\sigma =d=1$, which uses the Bohm-de Broglie
interpretation with the wave function $\Psi _1$, Equation (\ref{psi1}).
Each arrow of the vector field is shaded according to its true length, black
representing short vectors and white, long ones.
The four shades of gray show the regions where the vector field is pointing
to northeast, northwest, southeast or southwest. The black curves are the 
nullcline curves that separate these regions.
The trajectories are the white curves with direction arrows.
}
\end{center}
\end{figure}

A field plot of this planar system is shown in Figure 1 for $\sigma
=d=1$. We can see plenty of different possibilities, depending on the
initial conditions. Near the center points we can have oscillating
universes without singularities and with periodic isotropic phases, with
amplitude of oscillation of order $1$. For negative values of $\beta
_{0}$, the universe arises classically from a singularity but quantum
effects become important forcing it to recollapse to another
singularity, recovering classical behaviour near it. Isotropic phases
may happen near their maximum size. For positive values of $\beta _{0}$,
the universe contracts classically but when $\beta _{0}$ is small enough
quantum effects become important creating an inflationary phase which
avoids the singularity. The universe contracts to a minimum size and
after reaching this point it expands forever, recovering the classical
limit when $\beta _{0}$ becomes sufficiently large. In this case
isotropic phases may happen when the universe is near its minimum size.
In both cases we see that isotropic phases happen only when quantum
effects are important. We can see that, for $\beta _{0}$ negative, we
have classical limit for small scale factor while for $\beta _{0}$
positive we have classical limit for big scale factor. In these models with
$d=1$, the isotropic phases
last very shortly. Only for $\mid d\mid <<1$ can the isotropic phases be
arbitrarily large because, as said above, the separation of the singular
points goes like $\pi/d$.

For the wave function $\Psi _{2}$, the analysis goes in the same way but
we have to interchange $\beta _{0}$ with $\beta _{+}$ in Figure 1. In 
this case, we
have also periodic solutions but the others are anisotropic universes
arising classically from a singularity, experiencing quantum effects in
the middle of their expansion when they become approximately isotropic,
and recovering their classical behaviour for large values of $\beta
_{0}$. Depending on the initial conditions, the isotropic phase can be
arbitrarily large.
There are no further possibilities.

\subsection{The scalar field model}

Let us study the spherical wave solutions (\ref{psi3}) of Equation
(\ref{wdw}). The guidance relations (\ref{gui}) are

\begin{equation}  \label{gui0'}
p_{0} = \partial _0 S = \Im \biggl(\frac{\partial _0 \Psi _3}{\Psi _3}%
\biggr) = -{\dot{y}}^{0} \;,
\end{equation}
\begin{equation}  \label{guii'}
p_{i} = \partial _i S = \Im \biggl(\frac{\partial _i \Psi _3}{\Psi _3}%
\biggr) = {\dot{y}}^{i} \;,
\end{equation}
where $S$ is the phase of the wave function. In terms of $f$ and $g$ the
above equations read

\begin{equation} \label{gui0}
{\dot{y}}^{0}=-\Im \biggl(\frac{f^{\prime }(y^{0}+y)+g^{\prime }(y^{0}-y)%
}{f(y^{0}+y)+g(y^{0}-y)}\biggr) \;,
\end{equation}
\begin{equation} \label{guii}
{\dot{y}}^{i}=\frac{y^{i}}{y} \Im \biggl(\frac{f^{\prime
}(y^{0}+y)-g^{\prime }(y^{0}-y)}{f(y^{0}+y)+g(y^{0}-y)}\biggr) \;,
\end{equation}
where the prime means derivative with respect to the argument of the
functions $f$ and $g$, and $\Im(z)$ is the imaginary part of the complex
number $z$.

From Equations (\ref{guii}) we obtain that

\begin{equation}  \label{yi}
\frac{dy^{i}}{dy^{j}} = \frac{y^i}{y^j} \;,
\end{equation}
which implies that $y^{i}(t)=c_{j}^{i}y^{j}(t)$, with no sum in $j$, where
the $c_{j}^{i}$ are real constants, $c_{j}^{i}=1/c_{i}^{j}$, and $%
c_{1}^{1}=c_{2}^{2}=c_{3}^{3}=1$. Hence, apart some positive multiplicative
constant, knowing about one of the $y^{i}$ means knowing about all $y^{i}$.
Consequently, we can reduce the four equations (\ref{gui0}) and (\ref{guii})
to a planar system by writing $y=C|y^{3}|$, with $C>1$, and working only
with $y^{0}$ and $y^{3}$, say. The planar system now reads

\begin{equation} \label{gui0p}
{\dot{y}}^{0}=-\Im \biggl(\frac{f^{\prime }(y^{0}+C|y^{3}|)+g^{\prime
}(y^{0}-C|y^{3}|)}{f(y^{0}+C|y^{3}|)+g(y^{0}-C|y^{3}|)}\biggr) \;,
\end{equation}
\begin{equation} \label{guiip}
{\dot{y}}^{3}=\frac{{\rm sign}(y^3)}{C} \Im \biggl(\frac{f^{\prime
}(y^{0}+C|y^{3}|)-g^{\prime }(y^{0}-C|y^{3}|)}{%
f(y^{0}+C|y^{3}|)+g(y^{0}-C|y^{3}|)}\biggr) \;.
\end{equation}
Note that if $f=g$, $y^{3}$ stabilizes at $y^{3}=0$ because ${\dot{y}}^{3}$
as well as all other time derivatives of $y^{3}$ are zero at this line. As $%
y^{i}(t)=c_{j}^{i}y^{j}(t)$, all $y^{i}(t)$ become zero, and the
cosmological model isotropizes forever once $y^{3}$ reaches this line. Of
course one can find solutions where $y^{3}$ never reaches this line, but in
this case there must be some region where ${\dot{y}}^{3}=0$, which implies ${%
\dot{y}}^{i}=0$, and this is an isotropic region. Consequently, quantum
anisotropic cosmological models with $f=g$ always have an isotropic phase,
which can become permanent in many cases.

As a concrete example, let us take the gaussian $\Psi _4$ given in Equation
(\ref{psi4}). It is a spherical wave solution of the Wheeler-DeWitt
equation (\ref{wdw}) with $f=-g$, and hence it does not necessarily have
isotropic phases as described above for the case $f=g$. Figure 2 shows a
field plot of this system for $d/\sigma ^{2}=-10^{-4}$ and $C=2$. In the
region $|\phi |>>|\beta _{0}|$ we have periods of isotropic expansion
because $\phi={\rm const.}$ implies $\beta _{\pm }={\rm const.}$.
Depending on the value of $|\phi |$, this isotropic phase can be
arbitrarily long. These trajectories are periodic and without
singularities. Due to quantum effects, no trajectory crosses the point
$\beta _{0}=\phi =0$. Figure 3 shows the bohmian trajectories for the wave
function given in Equation (\ref{psi4}), now with $d/\sigma ^{2}=10^{-4}$ and
$C=2$. They have some qualitative behaviours different from the precedent
case. The isotropic phases, now in contraction, are no longer periodic. They
are parts of universes that contract anisotropically from infinity, experience
a period of isotropy, and contract anisotropically to a singularity. Figure 3
shows that no trajectories crosses the point $\beta _{0}=\phi =0$.

In order to get some analytical insight over Figures 2 and 3,
we present the planar system obtained from the guidance relations
corresponding to the wave function (\ref{psi4}) in the approximation
(\ref{fapprox}):
\begin{equation}
{\dot{\beta}}_{0}=\frac{\frac{2d}{\sigma ^{2}}(3\beta _{0}^{4}+6C^{2}\beta
_{0}^{2}\phi ^{2}-C^{4}\phi ^{4})}{\beta _{0}^{2}(\beta _{0}^{2}-C^{2}\phi
^{2})^{2}+\frac{4d^{2}}{\sigma ^{4}}(3\beta _{0}^{2}+C^{2}\phi ^{2})^{2}} \;,
\end{equation}
\begin{equation}
{\dot{\phi}}=\frac{\frac{16d}{\sigma ^{2}}\beta _{0}^{3}\phi }{\beta
_{0}^{2}(\beta _{0}^{2}-C^{2}\phi ^{2})^{2}+\frac{4d^{2}}{\sigma ^{4}}%
(3\beta _{0}^{2}+C^{2}\phi ^{2})^{2}} \;,
\end{equation}
where we have reset $y^{0}=\beta _{0}$ and $y^{3}=\phi$. This approximation
is not reliable in the lines $\beta _{0}=\pm C\phi $. As one can see
immediately from these equations, $\beta _{\pm }={\rm const.}$ whenever
$|\phi |>>|\beta _{0}|$, and the sign of $d$ defines the trajectories
direction.

\begin{figure}
\begin{center}
\includegraphics[trim=0in -0.1in 0in 0in]{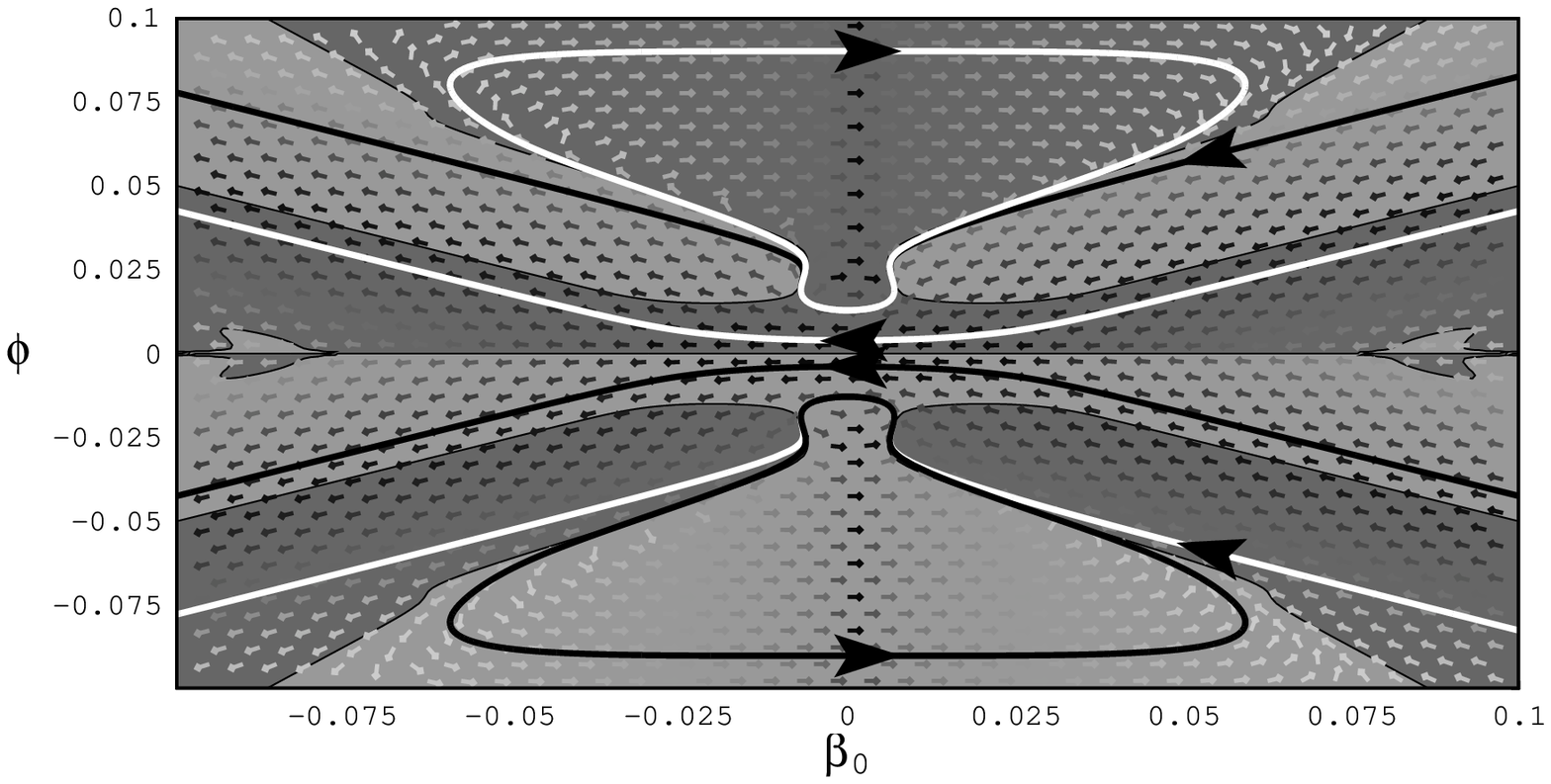}
\caption{
Field plot of the system of planar equations (\ref{gui0p}-\ref{guiip})
for $\sigma =100$, $d=-1$ and $C=2$, which uses the Bohm-de Broglie
interpretation with the wave function $\Psi _4$ given in Equation
(\ref{psi4}). Each arrow of the vector field is shaded according to its
true length, black representing short vectors and white, long ones. The
dark shade of gray shows the regions where the derivative of the vector
field points clockwise (the light shade of gray means the opposite), and
this shading allows to see the regions with arbitrarily long periods of
isotropic expansion corresponding to periodic universes without
singularities. There are also black or white curves with direction
arrows corresponding to universes contracting from infinity to a
singularity. Due to quantum effects, the trajectories do not cross the
origin.
}
\end{center}
\end{figure}

\begin{figure}
\begin{center}
\includegraphics[trim=0in -0.1in 0in 0in]{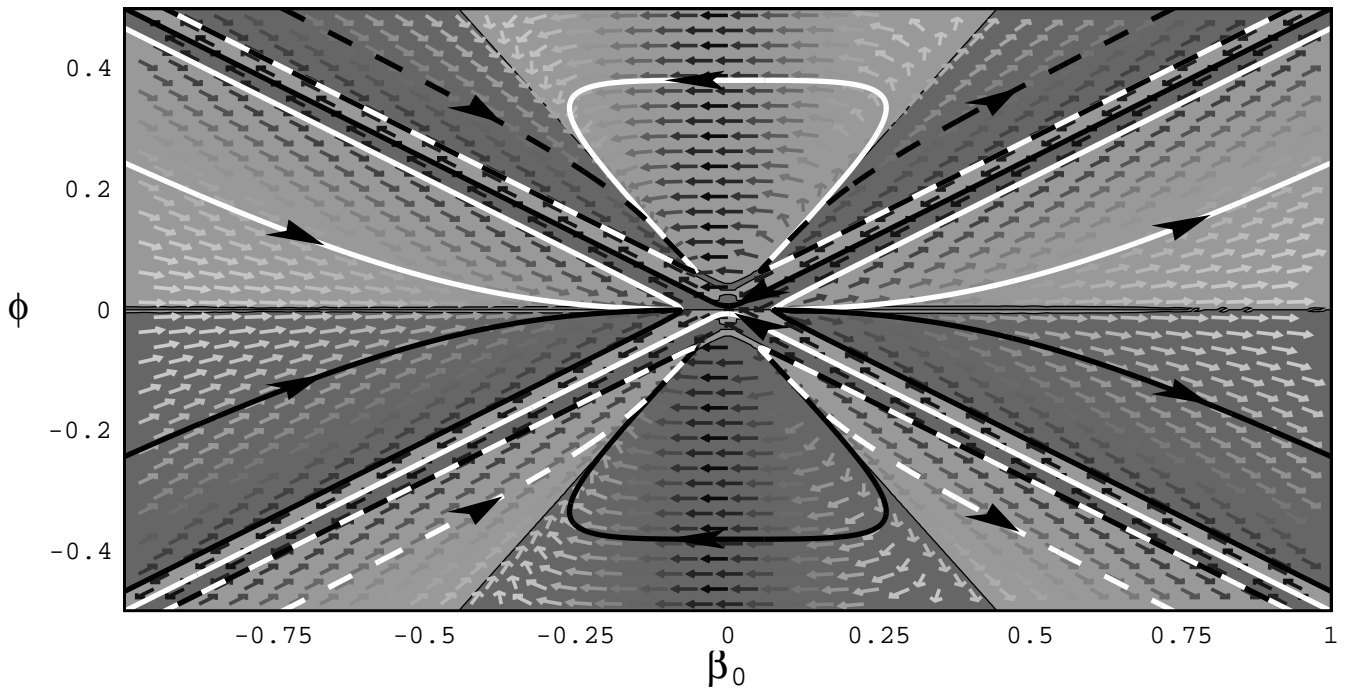}
\caption{
Field plot of the system of planar equations (\ref{gui0p}-\ref{guiip})
for $\sigma =100$, $d=1$ and $C=2$, which uses the Bohm-de Broglie
interpretation with the wave function $\Psi _4$, Equation (\ref{psi4}).
Each arrow of the vector field is shaded according to its true length,
black representing short vectors and white, long ones. The dark shade of
gray shows the regions where the derivative of the vector field points
clockwise (the light shade of gray means the opposite) and this shading
allows to see the regions with arbitrarily long periods of isotropic
contraction. The trajectories are the black or white curves with
direction arrows; they do not cross the origin due to quantum effects.
}
\end{center}
\end{figure}

\section{Conclusion}

Adopting the Bohm-de Broglie interpretation of quantum cosmology, we have
shown that quantum effects can generate an efficient alternative mechanism
for the isotropization of cosmological models. Anisotropic classical models
which never isotropize may present arbitrarily large isotropic phases during
the course of their evolutions if quantum effects are taken into account,
without needing to introduce any classical inflationary phase. The models
studied were Bianchi I models in empty space or filled with a free massless
scalar field.

There are questions and developments which should be undertaken within this
approach: the dependence of the above results on boundary conditions of the
Wheeler-DeWitt equation, their generalizations to other Bianchi models, and
the conditions for which quantum effects can also induce homogeneity on
cosmological models which are classically inhomogeneous. The last issue is
by far the most interesting but also the most complicated one because the
implementation of the Bohm-de Broglie interpretation for inhomogeneous
cosmological models is much more subtle and involved \cite{nel}. These
questions will be the subject of our future investigations.

\section*{ACKNOWLEDGMENTS}

We would like to thank the Cosmology Group of CBPF for useful discussions,
and CNPq and CAPES of Brazil for financial support. One of us (NPN) would like
to thank the Laboratoire de Gravitation et Cosmologie Relativistes of
Universit\'e
Pierre et Marie Curie, where part of this work has been done, for hospitality.

\end{document}